\documentstyle[11pt,hrd_pasp,twoside,epsf]{article}
\begin{document}

\title{A new, cleaner colour-magnitude diagram for the metal-rich 
globular cluster NGC 6528 and the velocity dispersion in the Galactic Bulge}

\author{Sofia Feltzing}
\affil{Lund Observatory, Box 43, SE-221 00 Lund, Sweden}
\author{Rachel Johnson}
\affil{IoA, Cambridge, UK and ESO, Santiago, Chile}

\begin{abstract}
Using two epochs of observations with HST/WFPC2 we obtain the stellar
proper motions for all stars in the field. The proper motions are  used
to separate the bulge from the cluster stars. The stellar sequences in the
resulting colour-magnitude diagram (CMD) are better defined than in any
previously published CMD.
Using $\alpha$-enhanced stellar isochrones we find NGC 6528 to have a
probable age of $11\pm 2$ Gyrs, this is the first attempt to establish
the absolute age of NGC 6528. Previous studies have only
compared the fiducial ridge line for the cluster to that of other
globular clusters of similar metallicities.  Mainly the comparisons
have been with regards to NGC 6553 and 47 Tuc.  With the new
metallicity determinations for individual stars in both NGC 6553 and
NGC 6528 it is now clear that 47 Tuc  (at $-0.71$ dex) has a 
significantly lower metallicity than NGC 6528 and NGC 6553 and is
thus not a suitable comparison cluster as regards differential age
determinations.
A comparison with the fiducial line for NGC 6553 confirms results
in earlier studies, e.g. Ortolani et al. (1995), that the two clusters
indeed have very similar ages. 
From the measured velocities both the proper motion of the cluster
and the velocity dispersion in the Galactic bulge are found.
NGC 6528 is found to have a proper motion relative to the Galactic
bulge of  $<\mu_l>$ = 0.006 and $<\mu_b>$ = 0.044
arcsec per century. Using stars with $\sim 14 <V_{\rm 555} < 19$ (i.e. the
red giant branch and horizontal branch)
we find, for the Galactic bulge,  $\sigma_{l}= 0.33\pm 0.03$ 
and $\sigma_{b}=
0.25\pm0.02$ arcsec per century. This give $\sigma_l/\sigma_b=1.32\pm0.16$, 
consistent both with previous proper motion studies of K giants in the 
Galactic bulge as well as with predictions by models of the kinematics
of bulge stars.
\end{abstract}

\section{Introduction}

NGC 6528 is perhaps {\sl the} most metal-rich globular cluster known,
e.g. 
Carretta et al. (2001). It has also been used as a ``reference'' cluster in
studies of other clusters, e.g.  Davidge (2000).  
NGC 6528 is situated at $(l,b)=(1.14,-4.12)$, i.e. in the plane of the Galactic
disk and towards the Galactic bulge.  The first effect of this is that
it is heavily reddened by foreground dust, e.g.
Richtler et al. (1998), Heitsch \& Richtler (1999). Most recent
distance estimates put the cluster within less than 1 kpc from the
Galactic centre (e.g. Richtler et al. 1998). 
The close proximity to bulge stars further complicates the
interpretation of the colour-magnitude diagram (CMD). Since the bulge stars
and the cluster stars have roughly the same distance modulus
they are superimposed in the CMD. 
This effect has been noted to be particularly pronounced on
the red giant branch (Richtler et al. 1998).  

In a pre-study we noted in particular that the colour-magnitude
diagram of the globular cluster NGC 6528 closely resembled that of NGC
5927 {\sl if} the latter was superimposed on the colour-magnitude
diagram of Baade's window (for a set of representative
CMDs see Feltzing \& Gilmore 2000).  This is
consistent with the cluster being virtually inside the Galactic bulge
and thus having a large contribution of bulge stars in its
CMD. Moreover, if the cluster is as metal-rich as
indicated in previous studies (and now confirmed by Carretta et
al. 2001) then the red-giant branch as well as both the turn-off and
the horizontal branch of the bulge and globular cluster will appear at
virtually the same magnitudes and colours. The bulge stars will be
more spread out in the CMD than those in the
cluster, due to the large range of metallicities and ages present in
the bulge (e.g.  McWilliam \& Rich 1994, Feltzing \& Gilmore 2000).  
Thus the only way to obtain
a clean CMD for the cluster is to obtain
proper motions of the cluster stars relative to the bulge stars and
separate the two populations using their proper motions.

This conclusion prompted us to apply for HST time to obtain a 
second epoch of observations of NGC 6528 with WFPC2 in order to
derive the relative proper motion of the cluster as compared to
that of the Galactic bulge. 
We here report on the first results from this proper motion study.

\section{Data}

\begin{figure}
\plottwo{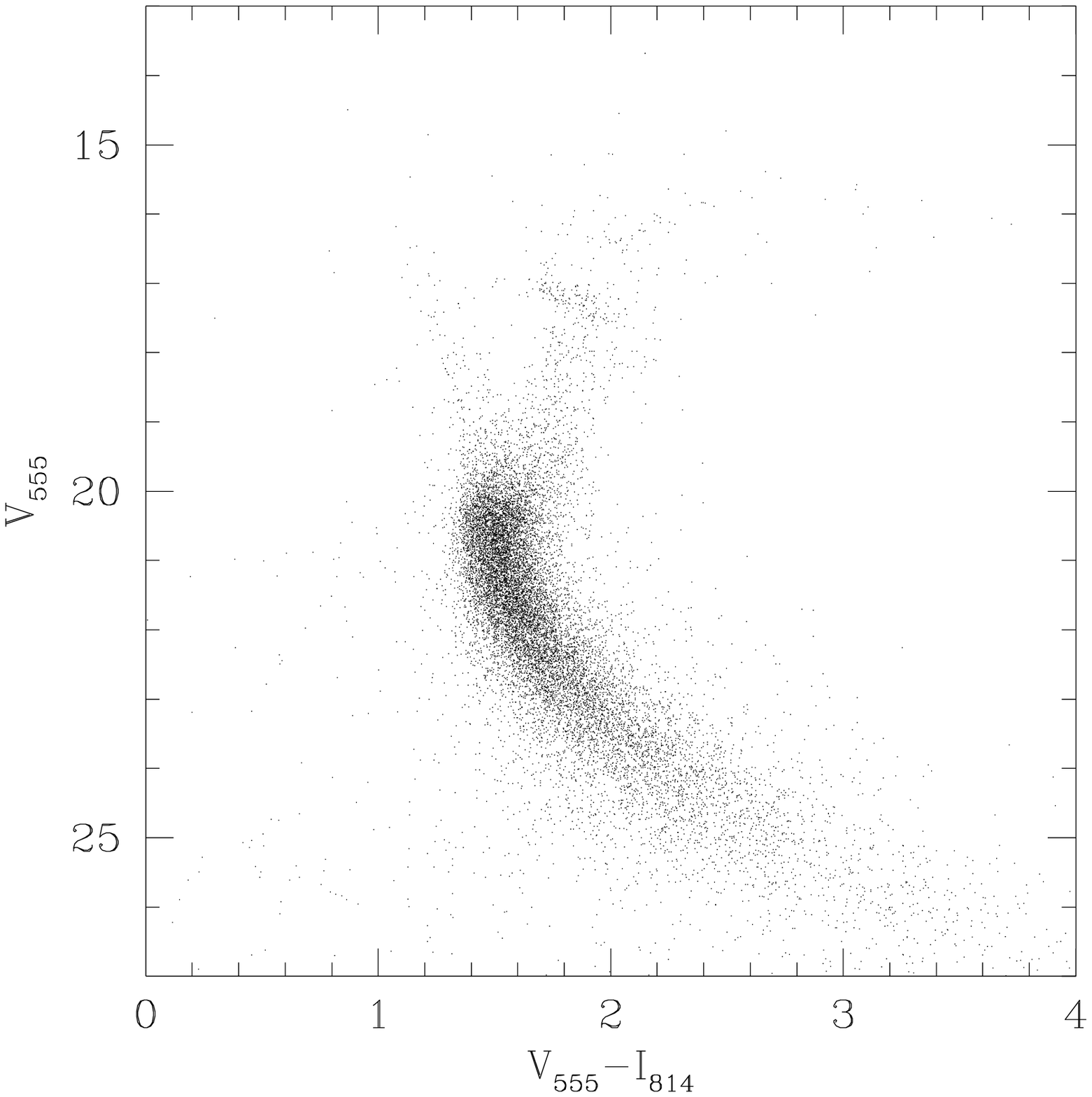}{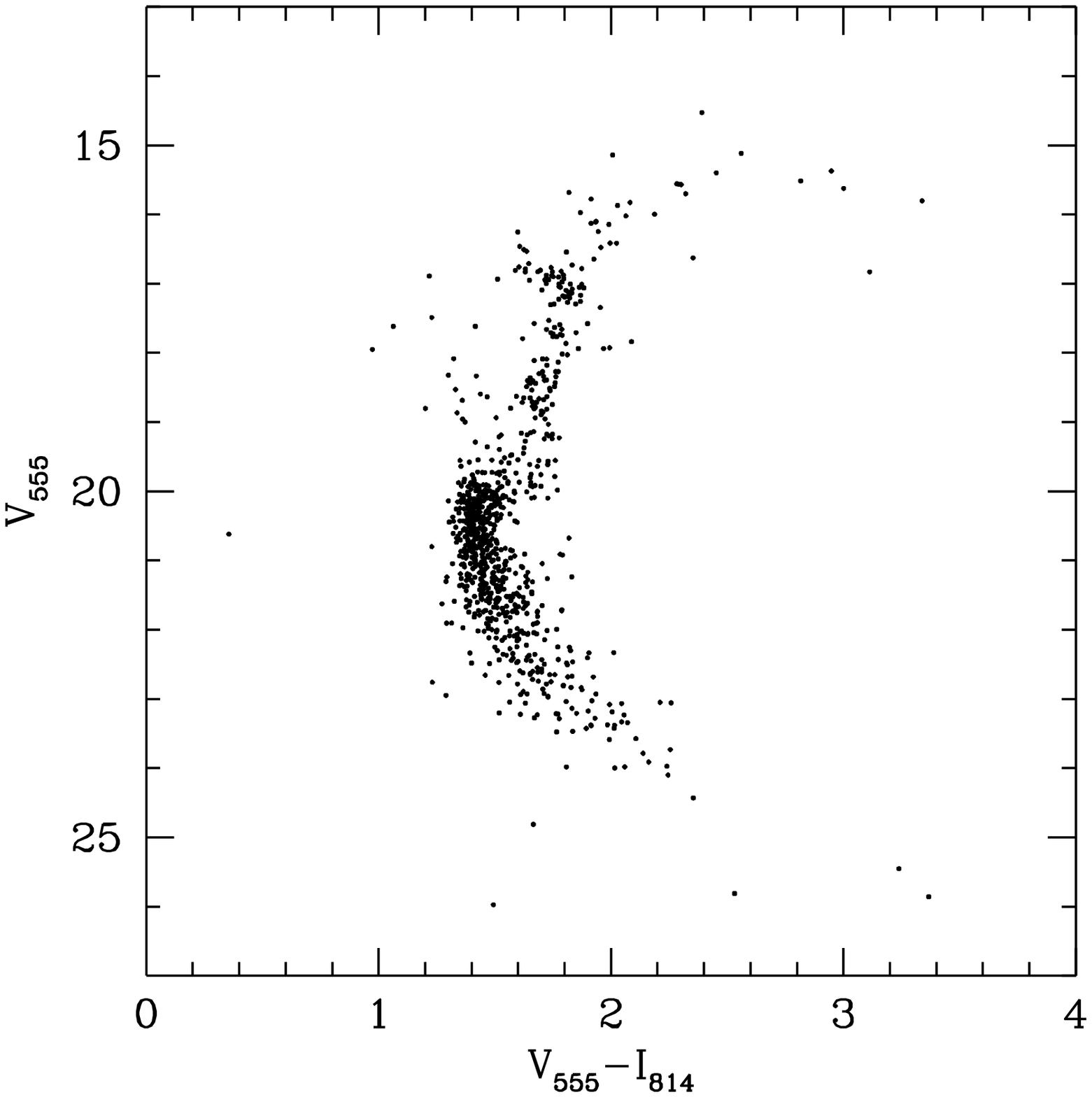}
\caption{{\bf Left panel:} The CMD for NGC 6528
based on all stars observed in the three WFs. {\bf Right panel:}
Final cleaned CMD. This data has also been corrected for 
differential reddening within the field.}
\end{figure}

The data consist of two sets of observations, one from the HST archive
and observed in 1994 (GO 5436) and the other our new data for the same
field (GO 8696, PI Feltzing).
Stellar photometry was done inside the {\sc digiphot.daophot}  package
in {\sc iraf} following standard routines, see e.g. Feltzing \& Gilmore (2000).
 Details of the photometry can be found
in Feltzing \& Johnson (2001, A\&A submitted). There we also describe
the routines we adopt to find the individual proper motions of the stars.

In an ideal scenario, where the globular cluster has an
appreciable motion in relation to the bulge stellar population, the
bulge and cluster stars will form two distinct distributions in the
proper motion diagram.  See for example
the recent results by Zoccali et al. (2001) for NGC 6553, King et
al. (1998) for NGC 6739 or Bedin et al. (2001) for M4
for good illustrations of this effect.

In our case the proper motions shows that the motion of the
globular cluster in relation to the Galactic bulge is very small.
This is as expected since the heliocentric radial velocity relative to the local
standard of rest for NGC 6528 is high ($184.9\pm3.8$ km s$^{-1}$
(Harris 1996), $\simeq$210 km s$^{-1}$ (Carretta et al. 2001)),
which suggests that NGC 6528 is on a mostly radial orbit away from us.
However, we note that the histograms for the velocities in the $l$ and $b$
coordinates have broad wings.
To separate the bulge and cluster stars using the measured proper
motions, we divide the stars into different magnitude ranges
and find the best fitting Gaussians to the proper motion histograms. 
 We found that two
Gaussians were required to fit the data well, indicating that, as
expected, we have (at least) two stellar populations with different velocity
dispersions.  Based on previous measurements of bulge and cluster
velocity dispersions, we associate the narrow Gaussian with NGC 6528
and the broad Gaussian with the bulge stars.  (For a full discussion 
of the analysis of 
the proper motion histograms the reader is referred to Feltzing \& Johnson
2001 submitted). 

\section{The cleaned CMD}

\begin{figure}
\plottwo{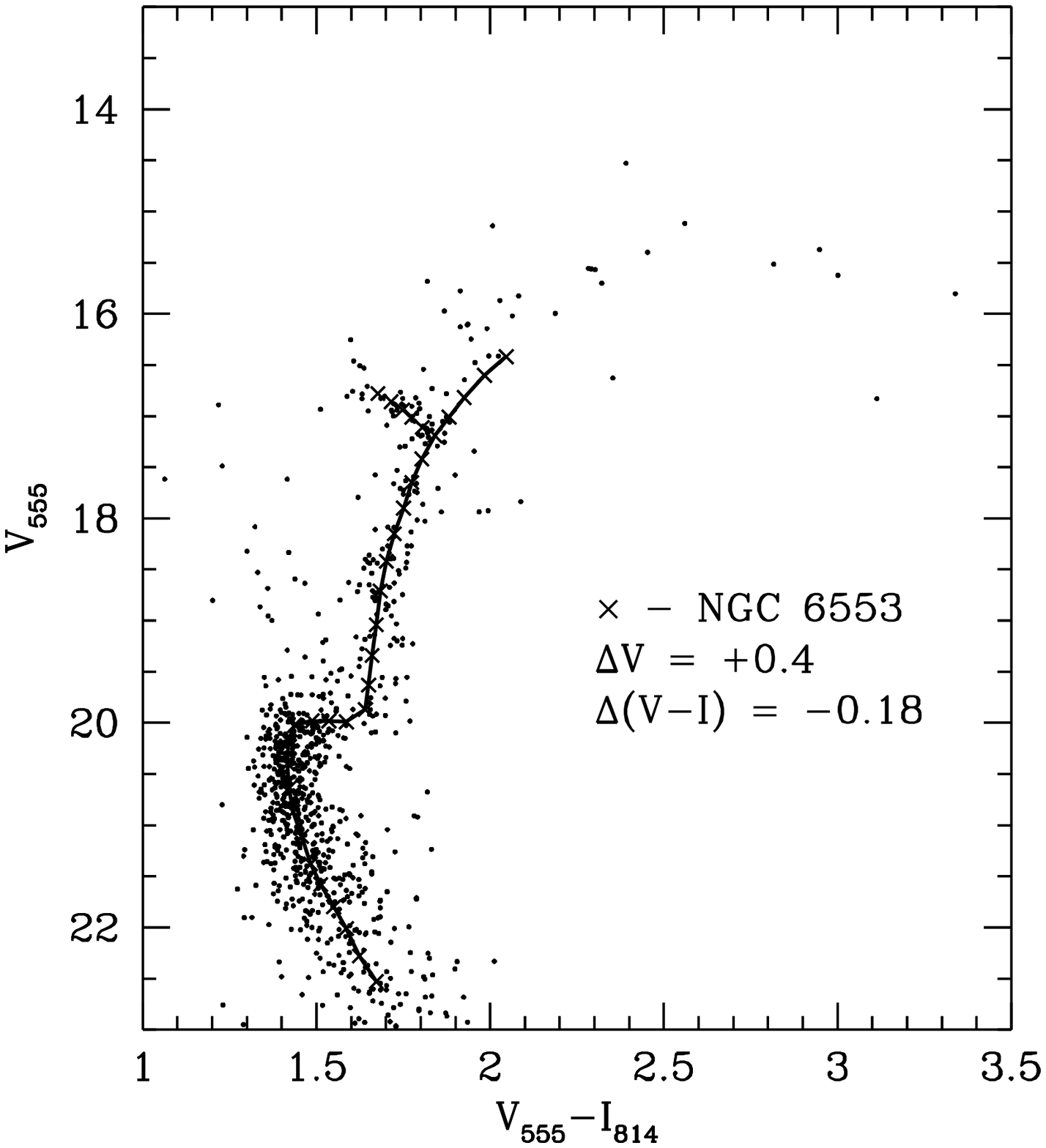}{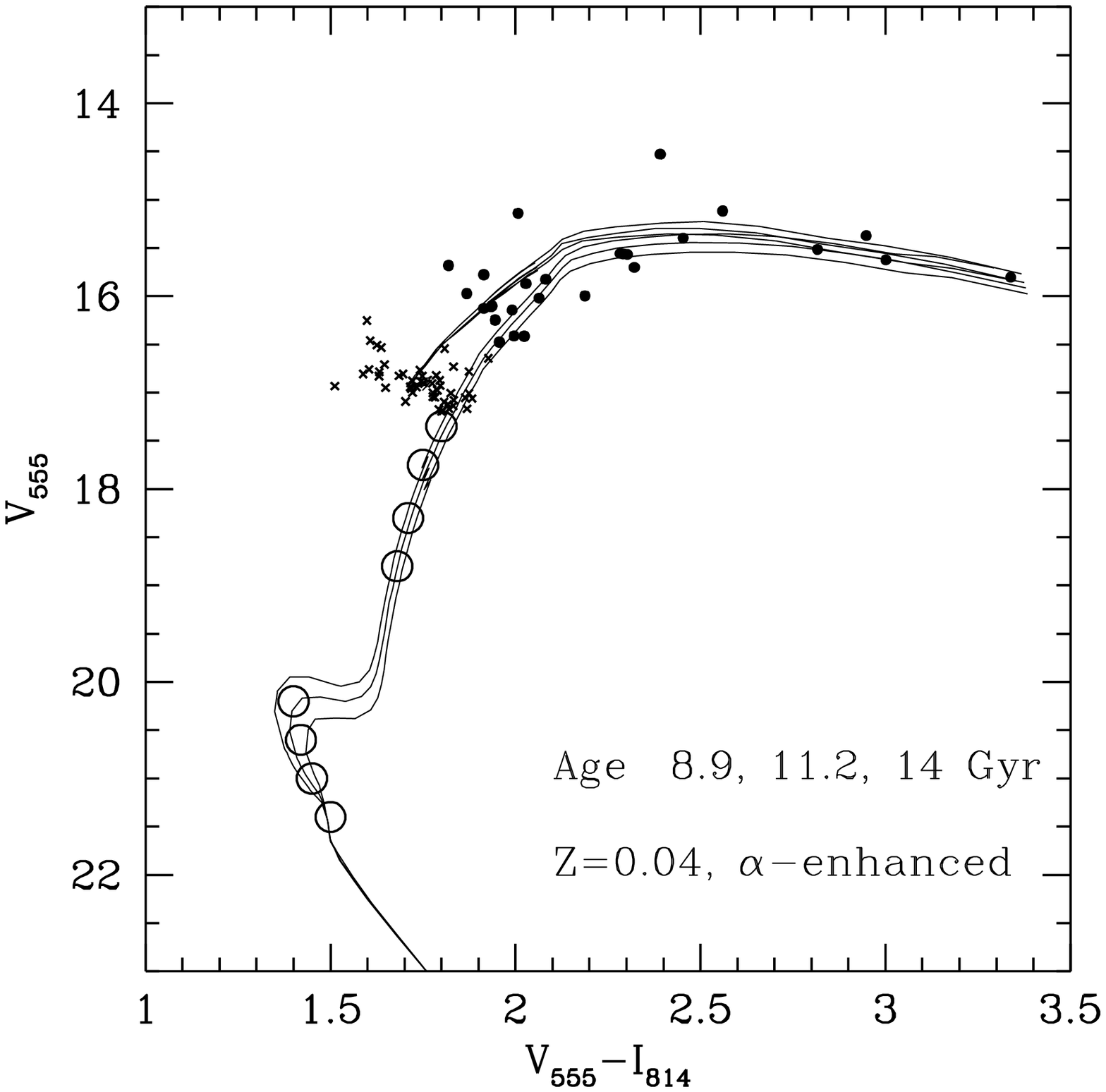}
\caption{{\bf Left panel:} The CMD for NGC 6528 with 
ridge line for NGC 6553 from Zoccali et al. (2001) over plotted. 
{\bf Right panel:}
Fiducial points for NGC 6528 main-sequence and stars on the RGB together
with the data for AGB and HB. $\alpha$-enhanced isochrones from 
Salasnich et al. (2000).  }
\end{figure}

We use our Gaussian fits to the different regions of the
CMD to estimate the proper motion cuts which
maximize the number of cluster stars relative to the number of bulge
stars, whilst still allowing enough cluster stars to make a good
cluster CMD. Using different proper motion cuts
in different regions of the CMD will affect the
relative numbers of cluster stars in each region but this does not
matter for comparison of the observed CMD with
other globular clusters and with isochrones, where all we use is the
position of the cluster stars in the CMD and not
their number density.

Our cleaned CMD, Fig. 1, is
finally obtained by imposing the following cuts;
$\sqrt{\mu_l^2+\mu_b^2}<0.18 $,  for star with $V_{\rm
555}<19$ and $\sqrt{\mu_l^2+\mu_b^2}<0.09 $ for the fainter stars.
We also find some differential reddening over the field of view of
WFPC2. Our final CMD is corrected for differential reddening.

\section{Age
from fitting stellar isochrones}

Since NGC 6528 is found to be enhanced, at least  in some,
$\alpha-$elements we  compare our fiducial points with that of
theoretical stellar isochrones from Salasnich et al. (2000) in which
$\alpha$-enhancement has been included. To facilitate the  comparison
with the stellar isochrones we define a set of fiducial points.

In the case of $Z=0.019$ we moved the isochrones by $\Delta(V_{\rm
555})=15.95$ and $\Delta(V_{\rm 555}-I_{\rm 814})=+0.63$. The turn-off
is well represented by the 11 Gyr isochrone and the horizontal branch
is well matched too. However, all the isochrones are 
brighter than the AGB. In order for $Z=0.019$ isochrones to
fit our data on the AGB we would need to increase the distance
modulus and the best fitting isochrone would then be very young,
younger than 9 Gyr. The horizontal branch would not be well fitted
either. Thus it appears unlikely that our data could be well fitted
with $\alpha$-enhanced isochrones with $Z=0.019$.

For the $Z=0.40$ isochrones we moved them with the following amount
$\Delta(V_{\rm 555})=15.95$ and  $\Delta(V_{\rm 555}-I_{\rm
814})=+0.655$, see 
Fig. 2. Here the AGB is much better reproduced and both turn-off
and horizontal branch can be well fitted simultaneously. The 11 Gyr
isochrone appears to fit best. However as this fit cannot be rigorous 
due to the limitations in the data the estimated error bar on this must
be rather large, perhaps up to 2 Gyr.

\section{The velocity dispersion in the bulge }

From the proper motions and the fitting of the proper motion
histograms we find that 
the velocity dispersion for the cluster is 0.08 arcsec per century.
This 
translates to 24-30 km s$^{-1}$ for the upper and lower distance limits
to NGC6528 of 6.5 kpc and 8 kpc (Richtler et al. 1998).

Globular clusters in the Galaxy have measured velocity dispersions
that range from a few km s$^{-1}$ to $\sim 20$ km s$^{-1}$, see Pryor \& Meylan
(1993) and Dubath et al. (1997). In M 31 at least two globular
clusters have measured velocity dispersions $>20$ km s$^{-1}$, Dubath \&
Grillmar (1997). Zoccali et al. (2001) found $\sigma$=28 km s$^{-1}$ for
NGC6553.  This result is very similar to ours. Since most globular
clusters in the Galaxy have significantly lower velocity dispersion
they concluded that their measured $\sigma_{cluster}$ was dominated by
measurement error. This is most likely also the case for NGC6528.

Assuming the cluster velocity dispersion in NGC 6528 is dominated by
errors, we deconvolve this from the measured velocity dispersion for
the bulge to find the true bulge velocity dispersion.  Using the data
for the bright red sample
 we get $\sigma_{l~bulge}= 0.33\pm 0.03$ and $\sigma_{b~bulge}=
0.25\pm0.02$ arcsec per century. These numbers are in good
agreement with the results for bulge giants found 
by Spaenhauer et al. (1992), $\sigma_{l~bulge}= 0.32\pm 0.01$ and $\sigma_{b~bulge}=
0.28\pm0.01$ arcsec per century for their full sample of 429 stars.

These numbers give a 
$\sigma_l/\sigma_b=1.32\pm0.16$, which is identical, within the error
estimates, to the 1.33 predicted for the coordinates of NGC 6528
 by the model of kinematics in the Galactic bulge
in Zhao (1996).

In their study of NGC 6553 Zoccali et al. (2001) derived
$\sigma_{l~bulge}= 0.26\pm 0.03$ and $\sigma_{b~bulge}= 0.21\pm0.02$
arcsec per century giving $\sigma_l/\sigma_b=1.24\pm0.17$. These
values are lower than found here, however, NGC 6553 is situated
further out from the Galactic centre than NGC 6528 and we should thus
expect $\sigma_{l~bulge}$ to be a factor  $\sim 0.86$ lower than for
the coordinates of NGC 6528, see Zhao (1996) Table 6.
$\sigma_{b~bulge}$ should remain roughly the same. Specifically the
model of Zhao (1996) predicts a $\sigma_l/\sigma_b=1.32$ at
$l=1,~b=-4$ and $\sigma_l/\sigma_b=1.09$  at $l=5,~b=-3$, which is
consistent, within the errors, to the values found here and in Zoccali
et al. (2001) for the bulge stars observed in the fields of NGC 6528
and NGC 6553 (which are situated close to the coordinates for which
Zhao's model makes its predictions).

We may thus conclude that these two new studies of the proper
motions of  Galactic bulge  stars confirm the predictions by models of
the kinematics in the  Galactic bulge. To our knowledge the current
work and that of Zoccali et al. (2001) are the first studies to
address the velocity dispersion, measured by proper motions, of bulge
stars below the horizontal branch.

\section{Summary}
CMDs of  the metal-rich globular cluster NGC 6528
are notoriously difficult to analyze. This is due to the fact the
cluster is situated in the Galactic bulge and thus the fields stars
belonging to the bulge have the same magnitudes as the stars in NGC
6528.

Using two epochs of observations with HST/WFPC2 we obtain the stellar
proper motions for all stars in the field. The proper motions are  used
to separate the bulge from the cluster stars. The stellar sequences in the
resulting CMD are better defined than in any
previously published CMD.

Using $\alpha$-enhanced stellar isochrones we find NGC 6528 to have a
probable age of $11\pm 2$ Gyrs, this is the first attempt to establish
the absolute age of NGC 6528. Previous studies have only
compared the fiducial ridge line for the cluster to that of other
globular clusters of similar metallicities.  Mainly the comparisons
have been with regards to NGC 6553 and 47 Tuc.  With the new
metallicity determinations for individual stars in both NGC 6553 and
NGC 6528 it is now clear that 47 Tuc  (at $-0.71$ dex) has a 
significantly lower metallicity than NGC 6528 and NGC 6553 and is
thus not a suitable comparison cluster as regards differential age
determinations (see e.g. Stetson et al. 1996 and references therein).

A comparison with the fiducial line for NGC 6553 confirms results
in earlier studies, e.g. Ortolani et al. (1995), that the two clusters
indeed have very similar ages. 

The stellar proper motion also provide velocity dispersions for both
the cluster and field stars. The velocity dispersion of the cluster
is most likely dominated by measurement errors. The bulge dispersion can
thus be found by deconvolution. The resulting dispersions are consistent
with what has previously (Spaenhauer et al. 1992) been found for Bulge
giants. Moreover, combining our results with those by Zoccali et al.
(2001) we are able to confirm the difference in $\sigma_l$ at two 
positions in the bulge as predicted by the model in Zhao (1996).

To our knowledge, our and Zoccali et al.'s study are the first to 
address the proper motions amongst bulge horizontal branch and fainter 
stars.

\end{document}